\documentclass[12pt]{article}
\usepackage{amssymb}
\usepackage{amsmath}
\usepackage{color}
\usepackage[normalem]{ulem}
\usepackage{verbatim}
\usepackage[colorlinks,linkcolor=blue,citecolor=green]{hyperref}
\usepackage{graphicx}
\allowdisplaybreaks[3]

\title{\textbf{Solutions to the wave equation 
		for commuting flows of dispersionless PDEs} }

\author{Natale Manganaro$^1$, \quad Alessandra Rizzo$^1$, \quad Pierandrea Vergallo$^{2,3}$\\
	\\
	\small  $^1$ Department of Mathematical, Computer,\\
	\small Physical and Earth Sciences\\
	\small University of Messina, \\
	\small V.le F. Stagno D'Alcontres 31, I-98166 Messina, Italy\\
          \small E-mail: nmanganaro@unime.it, alerizzo@unime.it \\
	$^2$ \small Department of Mathematics ‘F. Enriques’,\\
	\small  University of Milano, \\
	\small Via C. Saldini 50, Milano 20133, Italy\\
	$^3$ \small Sez. Milano, Istituto Nazionale di Fisica Nucleare,\\
	\small Via Celoria 16, Milano 20133, Italy \\
          \small E-mail: pierandrea.vergallo@unime.it
}
\date{\today}

\begin{document}
	\maketitle

	\begin{abstract}
		Motivated by the viewpoint of integrable systems,  we study commuting flows of 2-component quasilinear equations, reducing to investigate the solutions of  the wave equation with non-constant speed. 
		In this paper,  we apply the reduction procedure of differential constraints to obtain a complete set of solutions of such an equation for some fixed velocities $a^2(u,v)$. As a result, we present some examples of Hamiltonian integrable systems (as the shallow water equations) with relative symmetries, conserved quantities and solutions.
		
\vspace{0.4cm}
	\end{abstract}

	\maketitle
	
	\section{Introduction}
	Along the years many mathematical methods have been proposed for determining exact solutions of partial differential equations (PDEs). Among the others, in 1964 Yanenko \cite{jan} proposed the method of differential constraints and applied it to gas dynamics. The main idea is to add to the equations under interest some further differential relations. These play the role of constraints because they select classes of particular exact solutions admitted by the overdetermined set of equations (the original ones along with the additional differential constraints). The method is developed on two steps: first the compatibility of such an overdetermined system is  solved,  then exact solutions of the full set of equations is determined (the interested reader  can refer to \cite{mel1}).  If the resulting overdetermined system is in involution \cite{fsy,sh, rsy}, such a method is particularly useful for first order hyperbolic systems.  Indeed,  an increasing number of significative results have been obtained in the field of nonlinear waves \cite{ms1, m, cm, cwm, msh, mmw, mr1, jmr, mr2}. 
	
	Within such a theoretical framework, here we consider the wave equation 
	\begin{equation}
		f_{vv}-a^2 (u,v) f_{uu}=0,  \label{wave}
	\end{equation}
	where $a(u,v)$ is the non-constant wave speed.  The research of exact solutions of  (\ref{wave}) in a closed form  is of interest both for a theoretical point of view and  for possible applications.  In particular, this equation generalizes the celebrated wave equation with constant speed, which represents a key model in continuum mechanics. \\
	Furthermore, as we will show in the next sections,  equation (\ref{wave}) plays a key role in the theory of integrable systems, in particular for the one-dimensional quasilinear homogeneous first order system
	\begin{equation}\label{sy1}
		u^i_t=V^i_j(u)u^j_x, \qquad i=1,2\dots , n.
	\end{equation}
	Specifically,  in (\ref{sy1}), known also as {\it hydrodynamic type system}, $u^i(x,t), i=1,2,\dots n$ are the field variables depending on the space variable $x$ and  the time variable $t$; $V^i_j$ is an affinor (or simply the  matrix coefficient depending on $\textbf{u}$) and $\textbf{u}=(u^1,\dots u^n)$ is the $n$-component vector of field variables $u^j$. In Sections $2$ and $4$, we discuss simmetry, conservation laws and the related integrability of systems (\ref{sy1}) in connection with the solutions of (\ref{wave}).
	Preliminarly,  we briefly recall some basic notions on Hamiltonian formalism for evolutionary systems, i.e.  systems of the form
	\begin{equation}\label{es}
		u^i_t=F^i(\textbf{u},\textbf{u}_\sigma),\qquad i=1,2,\dots ,n,
	\end{equation} where $\sigma$ is a multi-index indicating the number of derivations with respect to the independent variable $x$, and $F^i$ are smooth functions of the arguments. Let us consider a linear differential operator $A^{ij}=a^{ij\sigma}D_\sigma$ and define  for every pair of  functionals $f$, $g$
	\begin{equation}\label{aaa1}
		\{f,g\}_A=\int{\frac{\delta f}{\delta u^i}A^{ij}\frac{\delta g}{\delta u^j}\, dx}.
	\end{equation} 
	The operator $A^{ij}$ is said a \emph{Hamiltonian operator} if the associated bracket (\ref{aaa1}) is a Poisson bracket, i.e. $(\ref{aaa1})$ is skew-symmetric and satisfies the Jacobi identity (by definition, the bracket satifies the Leibniz rule). Then (\ref{es}) is a \emph{Hamiltonian} if there exist a Hamiltonian operator $A^{ij}$ and a functional $H=\int{h(x,\textbf{u},\textbf{u}_\sigma)\,dx}$ such that 
	\begin{equation*}
		u^i_t=F^i(\textbf{u},\textbf{u}_\sigma)=A^{ij}\frac{\delta H}{\delta u^j}, \qquad i=1,2,\dots ,n,
	\end{equation*}
	where $\delta/\delta u^j$ is the variational derivative. 
	
	For the specific case of evolutionary system of hydrodynamic type (\ref{sy1}), B.A. Dubrovin and S.I. Novikov \cite{DNop}  introduced a class of homogeneous operators of first order
	\begin{equation}\label{DNop}
		P^{ij}=g^{ij}\partial_x+\Gamma^{ij}_ku^k_x,
	\end{equation}
	where the adjective \emph{homogeneous} refers to the order of derivation of each term (here the Einstein notation is used). In \cite{DNop, DN}  the authors proved that in the non-degenerate case (i.e.  $\det g^{ij}\neq 0$), the operator (\ref{DNop}) is Hamiltonian if and only if $g_{ij}=(g^{ij})^{-1}$ is a flat metric and $\Gamma^j_{ik}=-g_{is}\Gamma^{sj}_k$ are Christoffel symbols compatible with $g$. Operators (\ref{DNop}) naturally arise in systems (\ref{sy1}) whenever the Hamiltonian density $h$ depends on the field variables $\textbf{u}$ only.  In this case, if a hydrodynamic type system is Hamiltonian in the sense of Dubrovin and Novikov we have 
	\begin{equation*}
		u^i_t=V^i_j(\textbf{u})u^j_x \qquad \text{where} \quad V^i_j=g^{is}\nabla_s\nabla_j \, h(\textbf{u}).
	\end{equation*}
In  Section \ref{sec1} we introduce the notion of commutative flows for hydrodynamic type systems with the Hamiltonian property and we prove that their integrability is strictly related to the wave equation (\ref{wave}). 
	
 Within such a theoretical framework, the main aim of this paper is { to find exact solutions of (\ref{wave}).  As pointed out also by Makridin and Pavlov in \cite{pavlov2},  complete solutions to \eqref{wave} are very important because they explicitly represent the infinitely-many symmetries and conservation laws of Hamiltonian systems of hydrodynamic type \eqref{sy1} in 2-components.}
	
	In particular, in Section \ref{sec1} for the convenience of the reader, we motivate our investigation by introducing the main concept of commuting flows for hydrodynamic-type systems. In Section \ref{sec2} we determine  for specific cases the solution to the wave equation (\ref{wave}) parameterized in terms of two arbitrary functions. In Section \ref{sec4}, using the results obtained in the previous section, we characterize classes of Hamiltonian systems which we prove to be integrable.  Finally we give some comments and remarks in Section \ref{sec5}.
	
	\section{Commuting flows in the theory of  integrable  PDEs}\label{sec1}
	In this section, in view of studying Hamiltonian systems in two components, we sketch some general results obtained for Hamiltonian integrable systems. The theory of integrable systems is mostly based on the concept of commuting flows. Indeed, having infinitely many commuting flows guarantees the integrability property. For dispersionless systems of PDEs, we briefly recall that two systems \begin{equation}\label{gens1}
		u^i_t=V^i_ju^j_x, \qquad u^i_y=A^i_ju^j_x,
	\end{equation}
	are said to commute if and only if their flows commute, i.e. $u_{ty}-u_{yt}=0$.  {In such a case, the vector field \begin{equation}\label{symgen}\varphi=A^i_j(\textbf{u})u^j_x\frac{\partial}{\partial u^i}\end{equation} is a symmetry for the system $u^i_t=V^i_ju^j_x$ and viceversa.} Now, as we mentioned in the introduction, in \cite{DNop} it was proved that the natural local Hamiltonian structure for quasilinear systems of homogeneous first order PDEs is given in terms of first-order homogeneous operators. Moreover, in the nondegenerate case (i.e. when the leading coefficient $g^{ij}$ is nonsingular), being $g$ a flat covariant metric, it is always possible to apply local changes of complex coordinates such that 
	\begin{equation}\label{conmet}
		g^{ij}=\delta^{i,n+1-j}=:\eta^{ij}.
	\end{equation}
	Consequently, (see Corollary 2.2 in \cite{DubDef}) any Hamiltonian system of dispersionless PDEs can be reduced to 
	\begin{equation}\label{234}
		u^i_t=\eta^{ij}\frac{\partial^2 h}{\partial u^j \partial u^k}u^k_x,
	\end{equation}
	where the Hamiltonian functional is of hydrodynamic type (i.e. the Hamiltonian density $h$ only depends on the field variables and not on their derivatives).  The commutativity of systems of type \eqref{234} with Hamiltonian functionals $H=\int{h(\textbf{u})\, dx},F=\int{f(\textbf{u})\, dx}$ is then equivalent to the condition $\{H,F\}_{{P}}=0,$ where ${P}=\eta^{ij}\partial_x$ is the canonical Poisson bracket. In coordinates, this condition reads 
	\begin{equation}\label{genwaeq}
		\frac{\partial^2 h}{\partial u^i \partial u^l}\eta^{lk}\frac{\partial^2 f}{\partial u^k \partial u^j}-\frac{\partial^2 h}{\partial u^j \partial u^l}\eta^{lk}\frac{\partial^2 f}{\partial u^k \partial u^i}=0
	\end{equation}
	As far as we know, there is no general solution to condition (\ref{genwaeq}) even in the case of a lower number of components. Therefore, in the following Subsection we will investigate this equation in the case $n=2$.

	\subsection{Commuting flows {in 2-components}}\label{comflo} 
	Let us now focus on 2 components Hamiltonian quasilinear systems , i.e. hydrodynamic type systems of the form
	
	\vspace{0.1cm }
	
	\begin{equation}\label{gens}
		\left(  \begin{array}{c}
			u\\v
		\end{array}\right)_t=\left(\begin{array}{cc}
			0&1\\1&0
		\end{array}\right)\partial_x\left(\begin{array}{c}
			h_u\\h_v
		\end{array}\right), \quad  \left(\begin{array}{c}
			u\\v
		\end{array}\right)_y=\left(\begin{array}{cc}
			0&1\\1&0
		\end{array}\right)\partial_x\left(\begin{array}{c}
			f_u\\f_v
		\end{array}\right)
	\end{equation}
	whose Hamiltonian functionals are
	\begin{equation*}
		H=\int{h(u,v)\, dx}\qquad \qquad   F=\int{f(u,v)\, dx}.
	\end{equation*}
	
	\vspace{0.2cm}
	
	{ Relevant examples of this type in physics are the shallow water equations in Eulerian and Lagrangian coordinates,  the gas dynamic systems and the Hamiltonian systems describing a nonlinear elastic medium.  Such systems will be studied in details with their respective Hamiltonian densities in Section $4$.}
	
	\vspace{0.2cm}
	
	Now, considering the two component case of $(\ref{genwaeq})$, the following result holds.
	
	\vspace{0.4cm}
	
	\noindent \textbf{Theorem 1.} \label{thm34} The Hamiltonian systems
	in (\ref{gens}) commute if and only if 
	\begin{equation}\label{wwave2}
		h_{uu}f_{vv}-h_{vv}f_{uu}=0
	\end{equation}

 Let us assume that $h_{uu}h_{vv}\neq 0$, indeed if $h(u,v)$ is linear in $u$ or $v$ the equation \eqref{wwave2} trivially solves with a corresponding linear $f(u,v)$. However, corresponding Hamiltonian systems  turn out to be non-diagonalizable (see Section \ref{sec4}).
	
	Now  by requiring $h_{uu}h_{vv}>0$ and setting $\dfrac{h_{vv}}{h_{uu}}=a^2 (u,v)$, from (\ref{wwave2}) we get 
	\begin{eqnarray}
		&&f_{vv}-a^2 (u,v) f_{uu}=0,  \label{fwave} \end{eqnarray}
	along with 
	\begin{eqnarray}
		&&h_{vv}-a^2 (u,v) h_{uu}=0  \label{hwave}.
	\end{eqnarray}
	Therefore, for a given wave speed $a^2(u,v)$, any pair of functions $f(u,v)$, $h(u,v)$ determined from (\ref{fwave}) and (\ref{hwave}) ensure that the Hamiltonian systems (\ref{gens}) commute.   Furthermore, we notice that every function $f(u,v)$ commuting with the Hamiltonian density $h(u,v)$ is a conserved quantity for the Hamiltonian system defined by $h(u,v)$.  Viceversa,  $h(u,v)$ is a conserved quantity for the Hamiltonian system characterized by the Hamiltonian density $f(u,v)$. Hence,  solutions of the wave equation (\ref{wave}) are  also relevant  in the theory of integrable systems. 
	
	Finally, we remark  that if $\bar{f}(u,v)$ is solution to (\ref{wave}), then $\bar{f}(u,v)+c_1v+c_2u+c_3uv+c_4$ with $c_i$ constants gives a trivial new solution.  Therefore, in the next Section, we look for solutions of (\ref{wave}) up to the linear one $f(u,v)=c_1v+c_2u+c_3uv+c_4$. 		We emphasise that a general solution of \eqref{wwave2} is impossible to find in explicit form for an arbitrary function $h(u,v)$. Then, in what follows, we apply the method of differential constraints in order to restrict $h(u,v)$ such that equation \eqref{fwave} is completely solved.
	
	\section{Exact solutions to the wave equation}\label{sec2}
	In this Section, within the theoretical framework of the Method of Differential Constraints, we look for solutions of (\ref{wave}) which satisfy also the nonlinear differential constraint
	\begin{equation}
		\Phi(u,v,f,f_v,f_u)=0. \label{con}
	\end{equation}
	In passing we notice that the relation (\ref{con}) is called an intermediate integral of (\ref{wave}) if all solutions of (\ref{con}) are also solutions of (\ref{wave}) (for further details see \cite{mel1}).
	
	In order to study the compatibility between (\ref{wave}) and (\ref{con}), setting $P=f_v$, $Q=f_u$, $S=f_{vv}$, $R=f_{uu}$ and $W=f_{vu}$, differentiation of (\ref{con}) along with (\ref{wave}) leads to
	\begin{eqnarray}
		&&S-a^2 (v,u) R=0 \nonumber \\
		&&\Phi_u+\Phi_f Q+\Phi_Q R+ \Phi_P W=0 \label{c1} \\
		&&\Phi_v+\Phi_f P+\Phi_Q W+ \Phi_P S=0 \nonumber
	\end{eqnarray}
	Since here we are interested in determing exact  solutions of (\ref{wave})  in terms of arbitrary functions, we require that the determinant of the matrix coefficients of the system (\ref{c1}) in the unknown $S$, $R$ and $W$ is zero. Thus, we are led to the conditions
	$$
	\Phi_Q-a\Phi_P=0 \quad \mbox{or} \quad \Phi_Q +a \Phi_P=0  
	$$
	whose integration gives
	\begin{equation}
		\Phi=P-\lambda Q -g(v,u,f) \label{phi} 
	\end{equation}
	where we set $\lambda=\pm a$, while $g$ is a function to be determined. In the following, without loss of generality,  we will consider the case $\lambda=a$. Of course a similar analysis can be carried on in the remaining case $\lambda=-a$. Therefore, taking (\ref{phi}) into account, from (\ref{c1}) we obtain the following compatibility conditions
	\begin{eqnarray}
		&&\lambda_f=0  \nonumber \\
		&&\lambda_v+\lambda \lambda_u +2\lambda g_f=0 \label{f1} \\
		&&g_v+\lambda g_u+gg_f=0 \nonumber 
	\end{eqnarray}
	After some simple algebra, integration of (\ref{f1}) gives
	\begin{eqnarray}
		&&\lambda=\frac{1}{\left( A(\eta) v+ B(\eta) \right)^2} \label{l1} \\
		&&g=\sqrt{\lambda}\left( A(\eta) f +C(\eta) \right) \label{l2}
	\end{eqnarray}
	where, if $A(\eta) \neq 0$, $\eta$ is given by
	\begin{equation}
		u=-\frac{1}{A(\eta) \left( A(\eta) v + B(\eta) \right)}+\eta  \label{l3}
	\end{equation}
	while, if $A(\eta)=0$,  $\eta$ is defined by
	\begin{equation}
		u=\frac{v}{B^2(\eta)}+ \eta \label{l4}
	\end{equation}
	In (\ref{l1}), (\ref{l2}) $A(\eta)$, $B(\eta)$ and $C(\eta)$ are unspecified functions. Therefore, owing to (\ref{con}) and (\ref{phi}), exact solutions of (\ref{wave}) are obtained by solving the first order semilinear equation
	\begin{equation}
		f_v-\lambda f_u= \sqrt{\lambda}  \left( A(\eta) f+C(\eta) \right). \label{eq}
	\end{equation}
	It is of relevant interest to notice that, since the integration of (\ref{eq}) is parameterized by one arbitrary function and moreover the function $C(\eta)$ is arbitrary, by solving (\ref{eq}) by the method of characteristics, an exact solution of (\ref{wave}) is obtained in terms of two arbitrary functions. Therefore we were able to prove the following theorem:
	
	\vspace{0.2cm}
	
	\noindent \textbf{Theorem 2.} If the function $a(v,u)$ assumes the form
	$$
	a(v,u)=\frac{1}{\left( A(\eta)v+B(\eta) \right)^2} 
	$$
	with $\eta$ given by (\ref{l3}) or (\ref{l4}), then the solution of (\ref{wave}) can be obtained in terms of two arbitrary functions by integrating the first order equation (\ref{eq}). 
	
	\vspace{0.3cm}
	
	According to Theorem $2$, the wave speed $a^2 (u,v)$ is defined in implicit form by (\ref{l3}) or (\ref{l4}) so that, once $a(u,v)$ is given, the general solution of (\ref{wave}) will be determined by integrating (\ref{eq}). Hereafter, in order to give some explicit forms of $a(u,v)$ which allow the solution of the wave equation (\ref{wave}) in a closed form, we will consider some different cases.

	\vspace{0.4cm}
	
	{\it Case $1$.} Here, in the case where $A(\eta) \neq 0$, we assume
	$$
	A=\frac{1}{\sqrt{c_0  \eta}}, \quad \quad B=\frac{v_0}{\sqrt{c_0 \eta}}
	$$
	where $v_0$ and $c_0$ are arbitrary constants. In such a case, from (\ref{l1}) and (\ref{l3}) we obtain
	\begin{equation}
		a^2(u,v)=\frac{ c_0^2 u^2}{\left(v+v_0\right)^2 \left(  v+v_1 \right)^2}, \quad \eta=\frac{u(v+v_0)}{v+v_1}  \label{s1}
	\end{equation}
	where we set $v_1=v_0 -c_0$. Moreover,  integration of (\ref{eq}) gives the following solution of the wave equation  (\ref{wave})
	\begin{equation} 
		f=\sqrt{u \left( v+v_0 \right)\left(  v+v_1 \right) } \, \left( \theta_1(\eta)+ \theta_2(\sigma) \right) \label{sol1}
	\end{equation}
	where $\theta_1(\eta)$ and $\theta_2(\sigma)$ are arbitrary functions, while 
	$$
	\sigma=\frac{u\left(v+v_1 \right)}{v+v_0}.
	$$
	
	\vspace{0.4cm}
	\noindent
	{\it Case $2$.} We require $A(\eta)=0$ and $B(\eta)=\dfrac{v_0}{\eta}$, where $v_0$ is an arbitrary constant. In such a case, from (\ref{l1}) and (\ref{l4}) we get
	\begin{equation}
		a^2=\frac{u^2}{(v+v_0)^2}, \quad \quad \eta=\frac{v_0 \, u}{v+v_0}  \label{c2}
	\end{equation}
	while integration of (\ref{eq}) gives
	\begin{equation}
		f=\sqrt{u(v+v_0)}  \, \, (\theta_1 (\eta) + \theta_2 (\sigma)) \label{ss2}
	\end{equation}
	where $\theta_1 (\eta)$, $\theta_2 (\sigma)$ are arbitrary functions and
	\begin{equation}
		\sigma=u(v+v_0). \label{si}
	\end{equation}
	
	\vspace{0.4cm}
	\noindent
	{\it Case $3$.} Here, we assume $a(u,v)=p(v) \,q(u)$. Therefore, owing to the analysis above, from (\ref{f1}) we find that $p(v)$ and $q(u)$ must satisfy the relations
	\begin{eqnarray}
		&&2 \frac{d}{dv}\left( \frac{p^\prime}{p}\right)=p^2 \left( \frac{p^\prime}{p^2 }\right)^2+ k_0 p^2 \label{p} \\
		&&2 \frac{q^{\prime \prime}}{q^\prime}- \frac{q^\prime}{q}=-\frac{k_0}{q q^\prime} \label{q}
	\end{eqnarray}
	where $k_0$ is an arbitrary constant. Equations (\ref{p}) and (\ref{q}) can be easily integrated when $k_0 =0$. In such a case we have
	\begin{equation}
		p(v)=\frac{c_0}{(v+v_0)^2}, \quad \quad q(u)=c_1 (u+u_0)^2 \label{pq}
	\end{equation}
	where $c_0$, $c_1$, $v_0$ and $u_0$ are constants, so that the wave speeds $a^2 (u,v)$ assumes the form
	\begin{equation}
		a^2 = a_{0}^{2} \left( \frac{u+u_0}{v+v_0}\right)^4 \label{a3}
	\end{equation}
	with $a_0=c_0 c_1$. Finally, by integrating  (\ref{eq}) we obtain
	\begin{equation}
		f=(u+u_0)(v+v_0) \left( \theta_1 (\eta) + \theta_2 (\sigma) \right) \label{s4}
	\end{equation}
	where $\theta_1 (\eta)$ and $\theta_2 (\sigma)$ are arbitrary functions, while
	\begin{equation}
		\eta=\frac{1}{u+u_0}+\frac{a_0}{v+v_0}, \quad \quad \sigma=\frac{1}{u+u_0}-\frac{a_0}{v+v_0}.   \label{ns}
	\end{equation}
	Finally, as particular cases, we  first consider $q=1$, so that the function $a(v)$ adopts the form
	\begin{equation}
		a^2(v)=\frac{k_0^2 }{v^4} \label{s21}
	\end{equation}
	with $k_0$ an arbitrary constant, while the solution of (\ref{wave}) is given by
	\begin{equation}
		f=v \left( \theta_1(\eta) + \theta_2(\sigma) \right) \label{sol2}
	\end{equation}
	where $\theta_1(\eta)$ and $\theta_2(\sigma)$ are arbitrary functions and
	$$
	\eta=u+\frac{k_0 }{v}, \quad \quad \sigma= u- \frac{k_0 }{v}.
	$$
	Next, if we require $p=1$, we soon get
	\begin{equation}
		a^2(u)=k_1^4 u^4   \label{s3}
	\end{equation}
	where $k_1$ is an arbitrary constant, while integration of (\ref{eq}) gives
	\begin{equation}
		f=u \left(\theta_1(\eta) + \theta_2(\sigma) \right) \label{sol3}
	\end{equation}
	with $\theta_1(\eta)$ and $\theta_2(\sigma)$ arbitrary functions, while
	$$
	\eta=v+\frac{1}{k_1^2 u}, \quad \quad \sigma=v-\frac{1}{k_1^2 u}.
	$$

	\vspace{0.4cm}
	\noindent  \textbf{Remark 1.} Notice that cases defined by $(\ref{s21})$ and $(\ref{s3})$ are equivalent.  Indeed,  considering the change of dependent variables $u\leftrightarrow v$ and renaming $k_0=\dfrac{1}{k_1^2}$,  we map the first case onto the second one.  Indeed,  after the chosen change of variables we obtain
	\begin{equation}
		f_{uu}-\frac{1}{k_1^4u^4} f_{vv} =0 \longleftrightarrow f_{vv}-k_1^4u^4f_{uu}=0.
	\end{equation}
	
	\vspace{0.4cm}
	
	\noindent \textbf{Remark 2.} The solutions of the wave equation (\ref{wave}) given in cases $1$, $2$, $3$, are obtained iff the coefficient $a(v,u)$ takes one of the form (\ref{s1})$_1$,(\ref{c2})$_1$, (\ref{a3}),  (\ref{s21}) or (\ref{s3})$_1$. If we want to determine solutions of (\ref{wave}) with arbitrary coefficient  $a(v,u)$ we need to adopt a different strategy.  For instance, in the case $a(v)$, looking for solution under the form 
	\begin{equation}
		f(u,v)=F(u)G(v),   \label{sol4}
	\end{equation}
	we find
	\begin{eqnarray}
		&&F^{\prime \prime} (u) - \hat{k} F(u)=0 \label{r1} \\
		&&G^{\prime \prime}(v)+\hat{k}^2 a^2 (v) G(v)=0 \label{r2}
	\end{eqnarray}
	where $\hat{k}$ is an arbitrary constant. Therefore, once the coefficient $a(v)$ is specified, from (\ref{r1}), (\ref{r2}) different exact particular solutions of (\ref{wave}) can be determined depending if the constant $\hat{k}$ is positive or negative. Of course similar results can be obtained in the case $a(u)$.
	
	\section{Examples}\label{sec4}
	In this section we characterize classes of { integrable Hamiltonian systems}   for which we can explicitly present infinite commuting flows.  The importance of finding commuting flows is double: for each commuting flow we explicitly find a symmetry of the system (as in \eqref{symgen}) and a corresponding conserved quantity. Indeed, let us remark that conservation laws for quasilinear systems are 1-forms $\omega=Adt +Bdx$ closed modulo the equation.  Then,  in the 2-component hydrodynamic case, i.e. $A=g(u,v), B=f(u,v)$, they satisfy
		\begin{equation}
			(f(u,v))_t=(g(u,v))_x,
		\end{equation} where $f(u,v)$ is solution of \eqref{wave}.  In this section we will present some explicit examples by using the results {\color{blue}of} Section $3$.		Such symmetries and conserved quantities could be of interest for their physical meaning,  which is not strictly object of this paper.

		Before listing some examples, let us recall that for hydrodynamic-type systems $u^i_t=V^i_ju^j_x$,  the Haantjes
tensor
\begin{equation}
  H^i_{jk}=N^i_{pr}V^p_jV^r_k-N^p_{jr}V^i_pV^r_k
  - N^p_{rk}V^i_pV^r_j + N^p_{jk}V^i_rV^r_p,
\end{equation}
where $N^i_{jk}$ is the Nijenhuis tensor
\begin{equation}\label{nij}
  N^i_{jk}=V^p_jV^i_{kp}-V^p_{k}V^i_{jp}-V^i_p(V^p_{kj}-V^p_{jk}),
\end{equation}
is strictly related with their diagonalizability (see \cite{haantjes55:_x}). In particular, if $H^i_{jk}=0$ and the eigenvalues of the velocity
matrix $V^i_{,j}$ have the same algebraic multiplicity as their geometric
multiplicity, then the system of PDEs is diagonalizable. For systems in 2 components, the annihilation of the Haantjes tensor is  trivially satisfied. Then, a hydrodynamic-type system is diagonalizable depending on the number of distinct eigenvalues and their related eigenspaces. For systems in form \eqref{234},  we notice that the matrix of velocities has the following form:
\begin{equation}
V^i_j=\begin{pmatrix}h_{uv}&h_{vv}\\h_{uu}&h_{uv}\end{pmatrix}.
\end{equation} For such systems,  two eigenvalues coincide if and only if the discriminant of the chacteristic polynomial annihilates, i.e. if $h_{uu}h_{vv}=0$. It follows that two cases arise: \begin{itemize}
\item[a.] The hamiltonian density is nonlinear in both the field variables, the corresponding equation for commuting flows reduces to the wave equation and all the corresponding systems are diagonalizable. For such systems, we apply what found in the previous Section in order to explicit all the commuting flows for additive and multiplicative separable examples with each of the studied cases; 
\item[b.] The hamiltonian density is linear in one of the field variables and the corresponding system could be non-diagonalizable. In this case, the wave equation trivially solves with a linear $f(u,v)$.  \end{itemize}

Let us consider the two cases in details. 

\vspace{0.5cm}

\noindent \textbf{ a. Distinct eigenvalues}

	In  Subsection $(2.1)$ we verified that Hamiltonian systems like (\ref{gens}) commute if their Hamiltonian densities satisfy (\ref{wwave2}).  In the previous Section we find solutions of the wave equation (\ref{wave}) in terms of two arbitrary functions. Hence,  any Hamiltonian density $h(u,v)$ arising from (\ref{hwave}) characterizes a Hamiltonian system admitting infinitely many commuting flows. These are determined by the infinitely many Hamiltonian densities $f(u,v)$ solutions of (\ref{fwave}), implying the integrability of the system.  In such a way, for a fixed wave speed $a^2(u,v)$ chosen according to (\ref{s1}), (\ref{s21}) or (\ref{s3}), we are able to  determine infinitely many Hamiltonian systems characterized by $h(u,v)$, which are integrable. In what follows, we will give some examples of densities $h(u,v)$.
	
	\vspace{0.3cm}
	\noindent
	{\it Case I)} First, we consider the case of a Hamiltonian density $h(u,v)$ in  the additively separable form 
	\begin{equation}
		h(u,v)=H_1(u)+H_2(v). \label{h1}
	\end{equation}
	Substituting (\ref{h1}) in (\ref{hwave}), we find the following densities.
	
	\vspace{0,2cm}
	\noindent
	{\it i)} If $a^2(u,v)$ is given by (\ref{s1}), then
	\begin{equation}
		h(u,v)=-c_1 \ln{u}+\frac{ c_0 c_1  \left(2v+v_0+v_1\right)}{\left(v_1-v_0\right)^3}\ln{\left(\frac{v+v_0}{v+v_1}\right)} \label{t1},
	\end{equation}
where $c_1$ is an arbitrary constant. The corresponding Hamiltonian system admits infinitely many commuting flows determined by $(\ref{sol1})$.
	
	\vspace{0,4cm}
	\noindent
	{\it ii)} If $a^2(u,v)$ is given by (\ref{c2}), we have
	\begin{equation}
		h(u,v)=c_1 \ln(u(v+ v_0)),
	\end{equation}
	where $c_1$ is an arbitrary constant.
	The corresponding Hamiltonian system admits infinitely many commuting flows determined by (\ref{ss2}).
	
	\vspace{0,4cm}
	\noindent
	{\it iii)} If $a^2(u,v)$ is given by (\ref{a3}), we have
	\begin{equation}
		h(u,v)= \dfrac{c_1}{6} \left[\dfrac{1}{a_0^2(u+u_0)^2}+\dfrac{1}{(v+v_0)^2} \right],
	\end{equation}
where $c_1$ is an arbitrary constant.	The corresponding Hamiltonian system admits infinitely many commuting flows determined by (\ref{s4}).
	
	\vspace{0,4cm}
	\noindent
	{\it iiii)} If $a^2(u,v)$ is given by (\ref{s21}), we have
	\begin{equation}
		h(u,v)=\frac{c_1 u^2}{2}+\frac{c_1 k_0^2}{6v^2}, \label{t2}
	\end{equation}
where $c_1$ is an arbitrary constant. The corresponding Hamiltonian system commutes with infinitely many Hamiltonian systems characterized by $f(u,v)$ given by (\ref{sol2}).

	\vspace{0.4cm}
	\noindent
	{\it Case II)} Here we  look for Hamiltonian densities $h(u,v)$ of the form
	\begin{equation}
		h(u,v)=H_1(u) H_2(v). \label{h2}
	\end{equation} 
	By inserting (\ref{h2}) in (\ref{hwave}), after some algebra, we find the following 
	
	\vspace{0,2cm}
	
	\noindent
	{\it i)} If $a^2(u,v)$ is given by (\ref{s1}), then
	\begin{subequations}
	\begin{eqnarray}
		&&H_1(u)=c_1 u^{\frac{c_0+ \sqrt{c_0^2+ 4 k_0}}{2 c_0}}+ c_2 u^{\frac{c_0- \sqrt{c_0^2+ 4 k_0}}{2 c_0}},\\
		&&H_2(v)= \sqrt{(v+v_0)(v+v_1)} \left( c_3 \left( \frac{v+v_1}{v+v_0} \right)^{k_1} + c_4 \left( \frac{v+v_0}{v+v_1} \right)^{k_1} \right),
	\end{eqnarray}\end{subequations}
	where
	\begin{equation*}
		k_1=\frac{\sqrt{(v_0-v_1)^2+4k_0}}{2(v_0-v_1)}
	\end{equation*}
and $c_1$, $c_2$, $c_3$, $c_4$ are arbitrary constants.
	
	\vspace{0.4cm}
	\noindent
	{\it ii)} If $a^2(u,v)$ is given by (\ref{c2}), we have 
	\begin{subequations}
	\begin{eqnarray}
		&&H_1(u)=c_1 u^{\frac{1}{2} (1- c_0)}+ c_2 u^{\frac{1}{2} (1+ c_0)},\\
		&&H_2(v)= c_3 (v+v_0)^{\frac{1}{2} (1- c_0)}+ c_4 (v+ v_0)^{\frac{1}{2} (1+ c_0)},
	\end{eqnarray}
	\end{subequations}
	where $c_0, c_1, c_2, c_3, c_4$ are arbitrary constants.
	
	\vspace{0.4cm}
	\noindent
	{\it iii)} If $a^2(u,v)$ is given by (\ref{a3}), we have 
	\begin{subequations}
		\begin{eqnarray}
		&&H_1(u)=(u+u_0)  \left(c_1 e^{\frac{\sqrt{k}}{a_0(u+u_0)}}+ c_2 e^{-\frac{\sqrt{k}}{a_0(u+u_0)}}\right) \\
		&&H_2(v)= (v+v_0)  	\left(c_3 e^{\frac{\sqrt{k}}{v+v_0}}+ c_4 e^{-\frac{\sqrt{k}}{v+v_0}} \right),
	\end{eqnarray}
	\end{subequations}
	for $k>0$ and 
	\begin{subequations}
		\begin{eqnarray}
		&&H_1(u)=(u+u_0)  \left(c_1 e^{\frac{\sqrt{-k}}{a_0(u+u_0)}}+ c_2 e^{-\frac{\sqrt{-k}}{a_0(u+u_0)}}\right) \\
		&&H_2(v)= (v+v_0)  	\left(c_3 e^{\frac{\sqrt{-k}}{v+v_0}}+ c_4 e^{-\frac{\sqrt{-k}}{v+v_0}} \right),
	\end{eqnarray}
	\end{subequations}
for $k<0$. Here, $c_1$, $c_2$, $c_3$, $c_4$ are arbitrary constants.

	\vspace{0.4cm}
	\noindent
	{\it iiii)} If $a^2(u,v)$ is given by (\ref{s21}), we have 
	\begin{subequations}
	\begin{eqnarray}
		&&H_1(u)=c_1 u \sinh\left( \frac{1}{k_0 u \sqrt{k_1}} \right)+c_2 u \cosh\left(\frac{1}{k_0 u \sqrt{k_1}}\right),\\
		&&H_2(v)=c_3 e^{ \dfrac{v}{\sqrt{k_1}}} +c_4 e^{- \dfrac{v}{\sqrt{k_1}}},
	\end{eqnarray}\\
	\end{subequations}
	for $k_1>0$ and 
	\begin{subequations}
	\begin{eqnarray}
		&&H_1(u)=c_1 u \sin\left( \frac{1}{k_0^2 u \sqrt{-k_1}} \right)+c_2 u \cos \left(\frac{1}{k_0^2 u \sqrt{-k_1}}\right),\\
		&&H_2(v)=c_3 \sin \left( \frac{v}{\sqrt{-k_1}} \right) +c_4 \cos \left( \frac{v}{\sqrt{-k_1}} \right),
	\end{eqnarray} for $k_1<0$.  Here, $c_1$, $c_2$, $c_3$, $c_4$ are arbitrary constants.
	\end{subequations}
	
	\vspace{0.4cm}
	
	\noindent
	{\bf Remark 3.}
	In the theory of dispersionless systems, we say that a Hamiltonian density is said \emph{separable}  if there exist two functions of one variable $\alpha(u),\beta(v)$ such that
	\begin{equation}\label{82}
		\frac{h_{vv}}{h_{uu}}=\frac{\alpha(u)}{\beta(v)}.
	\end{equation}
	Moreover, if $\alpha(u)=1$ we say that it is a \emph{generalized gas Hamiltonian density}. 
	
	\vspace{4mm}
	
	In \cite{olvernutku}, the authors investigate equation (\ref{wwave2}) in the separable case.  Starting from the trival solution ${f}^*$, they construct an infinite number of conserved quantities recursively:
	\begin{equation}
		H_n(u,v)=\sum^n_{i=1}F_i(u)G_{n-i}(v),
	\end{equation} 
	where $F$ and $G$ satisfy
	\begin{equation}\label{12781}
		\frac{\partial^2 F_i}{\partial u^2}=\alpha(u)F_{i-1}, \qquad \frac{\partial^2 G_i}{\partial v^2}=\beta(v)G_{i-1},
	\end{equation}
	while $F_0,G_0$ are opportunely chosen from the trivial solution.  
	
	However,  explicit solutions are given in the simplest case $\alpha=1$ for which the authors obtain only polynomial solutions in $u$.  Here,  owing to the analysis developed in Section $3$,  we were able to find a more general class of solutions of (\ref{wave}), nonlinear in the field variables.  Indeed,  both \emph{Case 2} and \emph{Case 3} are in agreement with (\ref{82}) where,  $\alpha=1$ or $\beta=1$, respectively  while in \emph{Case 1}  we have
	$$
	\alpha(u)=c_0^2 u^2, \quad \quad \beta(v)=(v+v_0)^2 (v-v_1)^2.
	$$

	{
		{\bf Example 1. }\emph{[Shallow water equation in Eulerian coordinates]} In \cite{olvernutku}, the authors consider the Hamiltonian density
		\begin{equation}
			h(u,v)=-\frac{1}{2}u^2 v-s(v), \label{o1}
		\end{equation}
		which generates the isoentropic gas dynamics system. In such a case $u$ denotes the velocity, $v$ the mass density and 
		$
		s^{\prime \prime}(v)=\sigma^\prime (v)/{v},
		$
		where $\sigma (v)$ is the pressure. It is easy to verify that the Hamiltonian (\ref{o1}) is  solution to (\ref{hwave}) with $a^2(u,v)$ given by (\ref{s21}) if the we adopt for the pressure $\rho (v)$ the Von-K\'arm\'an law
		$$
		\rho (v)=-\frac{k_0^2}{v}+p_0,
		$$
		where $p_0$ is a constant. Therefore, in such a case, the gas dynamics system associated with (\ref{o1}) admits infinitely many commuting flows given  by $f(u,v)=v \left( \theta_1(\eta) + \theta_2(\sigma) \right)$,
			where $\theta_1(\eta)$ and $\theta_2(\sigma)$ are arbitrary functions as in \eqref{sol2}.
		
		\vspace{4mm}

		 {\bf Example 2.} \emph{[Shallow water equation in Lagrangian coordinates]}  In the same paper, the authors also consider the Hamiltonian density
		\begin{equation}
			h(u,v)=\frac{u^2}{2}+F(v) \label{o2}
		\end{equation}
		which gives the Hamiltonian system describing a nonlinear elastic medium. In such a case $u$ is the velocity, $v$ the strain,  and { $
			F^\prime (v)=-p(v)
			$}
		where $p$ denotes the stress. The density (\ref{o2}) is a solution to (\ref{hwave}) supplemented by (\ref{s21}) if 
		\begin{equation}
			p(v)=\frac{k_1^2}{3v^3}.  \label{ss4}
		\end{equation}
		Therefore the nonlinear elastic system supplemented by (\ref{ss4}) and generated by the density (\ref{o2}) is integrable because it commutes with infinitely many flows determined by (\ref{sol2}).
		
		\vspace{3mm}
		
		\noindent {\bf Remark 4.} Note that exchanging $u$ with $v$ in the expressions (\ref{o1}) and (\ref{o2}) we obtain the well-known Hamiltonian densities for the shallow water equation, respectively in Eulerian and Lagrangian coordinates. 
	}

 \vspace{3mm}
 
 The previous example is part of a more general family of equations, that are included in the celebrated $p$-system. Such a system is mainly studied in the context of wave propagation problems  \cite{smoller} and admits the following hydrodynamic-type structure
	\begin{align}\label{1}\begin{split}
		\begin{cases} v_t -u_x=0  \\
		 u_t+(p(v))_x=0 \end{cases}\end{split}
	\end{align} 
	where $p(v)$ is a smooth function with $p^\prime(v) <0$ and $p^{\prime \prime}(v) >0$.
	Equations in (\ref{1})  describe different situations of physical interest. For instance, in the case of  isentropic fluid dynamics in Lagrangian coordinates,  $v$  denotes the specific volume, $u$ the Lagrangian velocity and  $p(v)$ the pressure. In what follows, we want to use our results in Section \ref{sec2} to find infinitely many solutions of the underlined class of systems.  Firstly, according to \cite{olvernutku}, we notice that every $p$-system is Hamiltonian with respect to a first-order homogeneous operator in canonical form, i.e. $P^{ij}=\eta^{ij}\partial_x$. Indeed,  (\ref{1}) can be re-written as
	\medskip
	\begin{equation*}\label{ps}
		\left(\begin{array}{c}
			u\\v
		\end{array}\right)_t=\left(\begin{array}{cc}
			0&1\\1&0
		\end{array}\right)\partial_x\left(\begin{array}{c}
			h_u\\h_v
		\end{array}\right),
	\qquad
		h(u,v)=\frac{u^2}{2}+F(v)  %\label{H3}
	\end{equation*}
	where $F'(v)=-p(v)$.

	Now, let us briefly recall that the hodograph method \cite{yan2} introduces an exchange of dependent and independent coordinates and allows to reduce each $2\times 2$ hydrodynamic-type systems to the following
	\begin{equation}\label{hods}
	\begin{cases}x_v=-V^1_1t_v+V^1_2t_u\\x_u=V^2_1t_v-V^2_2t_u
	\end{cases}, 
	\end{equation}
	such that to every solution $(u,v)$ of the given system corresponds  a solution $(x,t)$ of the previous and viceversa.  Then, let us finally note that solving system \eqref{hods} is equivalent to solve the commutativity conditions between $u^i_t=V^i_ju^j_x$ and $u^i_y=\varphi^i_ju^j_x$ after introducing the transformation
	\begin{equation}\label{tsaeq}
	\varphi^i_j=V^i_jt+\delta^i_j x,
	\end{equation}
	The interested reader can see \cite{tsarev} for further details. In particular, one can obtain solutions of hydrodynamic-type systems by looking for solutions of \eqref{tsaeq} where $\varphi^i_j$ are coefficients of their hydrodynamic symmetries.  As proved in \cite{tsarev}, such symmetries are infinitely many and parametrized by 2 arbitrary functions. For the particular case of the $p$-system, the hydrodynamic symmetries are \begin{equation}\label{hysp}
		\varphi=\left(f_v\, u_x-p' (v)f_u\, v_x\right)\frac{\partial}{\partial u}+\left(f_u\, u_x+f_v\, v_x\right)\frac{\partial}{\partial v},
	\end{equation} where $f(u,v)$ is solution to the wave equation
	\begin{equation}
		f_{vv}+p'(v)f_{uu}=0. \label{vave}
	\end{equation} 
	
	Then, for the $p$-system solving equations \eqref{tsaeq} gives $x=f_v (u,v)$ and $t=f_u (u,v)$,
		where $f(u,v)$ is solution to \eqref{vave}. Therefore, finding the general solution to (\ref{vave}) yields to infinitely many solutions of the $p$-system, depending on the choice of the two arbitrary functions.

	\vspace{0.3cm}
	
	\noindent {\bf Example 3.} { Let us consider the following $p$ function 
		\begin{equation}\label{p1}
			p'(v)=-\frac{k_0^2}{v^4}.
		\end{equation}
		Then,  keeping in mind (\ref{s21}),  solutions of this $p$-system are given implicitly by (\ref{sol2}).  As expected, the symmetries are parametrized by two arbitrary functions and since (\ref{sol2}) is the general integral of (\ref{vave}), here we find the complete set of implicit solutions of (\ref{1}) supplemented by (\ref{p1}) computable via the hodograph method.}

	 \vspace*{6mm}
	 
\noindent  	\textbf{b. Coincident eigenvalues}

 Let us now assume without loss of generality that $h_{uu}=0$ (the transformation that exchanges the field variables maps this case into $h_{vv}=0$). Then,
\begin{equation}
h(u,v)=A(v)u+B(v)
\end{equation} and the related systems admit the following form
\begin{equation}\label{jbs}\begin{cases}
u_t=A'(v)u_x+\left(A''(v)u+B''(v)\right)v_x\\
v_t=\qquad \qquad \qquad A'(v)v_x\end{cases}.
\end{equation}Systems of this kind are diagonalizable if and only if $A,B$ are linear in $v$, i.e. the hamiltonian density is 
\begin{equation}\label{trivial1}
		h^*(u,v)=c_1u+c_2v+c_3uv+c_4
	\end{equation}
	where $c_1,c_2,c_3$ and $c_4$ are arbitrary constants. This density is a trivial solution of the equation \eqref{wwave2} that leads to the  trivial linear system $u_t=c_3 u_x, v_t=c_3v_x$. Then, we assume $A,B$ to be not both linear.
Note that such systems are in Jordan-block form where $V_1^1=V_2^2$ do not depend on $u$. Systems of this form were  investigated by E.V. Ferapontov and one of us in \cite{verfer1}. In particular, in the paper the authors studied the linear degeneracy property for Jordan-block systems as a necessary condition to admit a Hamitonian structure. We recall that a hydrodynamic-type system is said to be \emph{linearly degenerate} if each eigenvalue $\lambda_i$ is constant along the corresponding eigendirection $X_i$, i.e the Lie derivative $\mathcal{L}_{X_i}\lambda_i$ is equal to zero. Systems \eqref{jbs} considered in this paper are all linearly degenerate because admit a canonical Dubrovin-Novikov Hamiltonian structure.  Moreover, systems \eqref{jbs} possess infinitely many Hamiltonian structures:
\begin{equation}\label{hamdeg}
\begin{pmatrix}
\xi_2 u +\xi_3&\xi_1\\\xi_1&0
\end{pmatrix}\partial_x+\begin{pmatrix}
\dfrac{\xi_2}{2}&0 \vspace{0.15cm}\\0&0
\end{pmatrix}u_x+\begin{pmatrix}
\dfrac{\xi_2'u+\xi_3'}{2}&-\dfrac{\xi_2}{2}+\xi_1' \vspace{0.15cm} \\ \dfrac{\xi_2}{2}&0
\end{pmatrix}v_x
.\end{equation} where $\xi_1,\xi_2,\xi_3$ are arbitrary functions depending on $v$.
The multi-Hamiltonian structure of these systems (together with the very simple structure of \eqref{jbs} with an independent subsystem $v_t=A'(v)v_x$) makes this case not of interest. However, if we set $\xi_1(v)=0$ in   \eqref{hamdeg}, we obtain a multi-Hamiltonian \emph{degenerate} structure, i.e. $\det(g^{ij})=0$. The corresponding Hamiltonian density has the following form
\begin{align}\begin{split}
h(u,v)&=\frac{2A'(v)}{\xi_2}u-2\int{\frac{B''(v)\xi_2^2-\xi_3'A'(v)\xi_2-2A''(v)\xi_2\xi_3+2A'(v)\xi_2'\xi_3}{\xi_2^3}\, dv}.\end{split}
\end{align} 
As a result, this suggests the necessity of studying degenerate structures for non-diagonalisable systems.

\newpage
	\section{Conclusions}\label{sec5}
	{

	\noindent Determining exact solutions of higher order PDEs is a hard task so that any mathematical approach leading to solve PDEs is of great interest not only from a theoretical point of view but also for possible applications. Among others, the method of the intermediate integrals is particularly useful for second order hyperbolic equations. It leads to look for solutions of the PDEs under interest which satisfy a further differential constraint that, in principle, could be nonlinear. As it happens for all the reduction procedures proposed in the literature, such an approach permits to determine classes of exact solutions of the equation under investigation by solving a first order PDEs. Therefore the solutions obtained are determined in terms of at least one arbitrary function. Furthermore, when in the PDEs at hand we have some material response functions not assigned, then such a procedure allows to characterize classes of PDEs for which the reduction procedure holds and consequently exact solutions can be determined.\\Such features of the method of intermediate integrals suggested us its use for determining exact solutions of equation (\ref{wave}). In fact the results here obtained permitted us to characterize classes of wave equations like (\ref{wave}) for which the solution in terms of two arbitrary functions is obtained so that any initial value problem can be solved.	The connection of this equation to integrable systems was firstly investigated by Nutku and Olver in \cite{olvernutku},  who showed some solutions under the above introduced separable hypothesis.  Later,  in \cite{pavlov2}  Makridin and Pavlov presented an infinite number of solutions which are polynomial in one of the two field variables for the more general case when $h(u,v)=A(u)v^2+B(u)v+C(u)$.  In our present paper,  we reconsider the separable case but finding a larger class of solutions which are general for fixed speed, i.e.  depending on two arbitrary functions of one variable (as expected  \cite{tsarev}).  
		
		Apart from its theoretical value,  such a result was used to study symmetries, conservation laws and the Hamiltonian structure of some systems  which admit infinitly many commuting flows.  Finally we also used  the results obtained in Section $3$ to find simmetries and conserved quantities of the famous homogeneous p-system, i.e.  the shallow water equation in Lagrangian coordinates.  
		
		It could be of relevant interest to develop a similar analysis for systems involving higher number of components.  A future perspective of this research is also to find other general solutions for the wave equation where $a^2(u,v)$ is not separable,  i.e.  in the theory of integrable systems, looking for non-separable Hamiltonian densities.  Furthermore,  it would be of interest to apply this  procedure also to the deformation of the Hamiltonian structures for non-homogeneous operators (see \cite{ver1,ver2} for an investigation of Hamiltonian non-homogeneous systems of hydrodynamic type).  }

\subsection*{Acknowledgements}
		The authors thank Maxim Pavlov for stimulating discussions and interesting remarks. We  also acknowledges the financial support of GNFM of the Istituto Nazionale di Alta Matematica and of PRIN 2017 \textquotedblleft Multiscale phenomena in Continuum
		Mechanics: singular limits, off-equilibrium and transitions\textquotedblright,
		project number 2017YBKNCE. PV's research was partially supported by the research project Mathematical Methods in Non-Linear Physics
		(MMNLP) and by the Commissione Scientifica Nazionale -- Gruppo 4 -- Fisica Teorica
		of the Istituto Nazionale di Fisica Nucleare (INFN). AR’s research was partially supported by INdAM-GNFM
		through “Progetto Giovani GNFM 2023” entitled “p-systems in elastodinamica
		non lineare”.

\end{document}